\documentclass[%
 reprint,
 amsmath,amssymb,
 aps,
]{revtex4-2}

\usepackage{graphicx}
\usepackage{dcolumn}
\usepackage{bm}
\usepackage{xcolor}
\usepackage{amsmath}
\usepackage{tabularx}

\begin{document}

\preprint{APS/123-QED}

\preprint{APS/123-QED}
\title{Photon Conversion in Thin-film Lithium Niobate Nanowaveguides: A Noise Analysis}

\author{Heng Fan}
\author{Zhaohui Ma}%
\author{Jiayang Chen}
\author{Zhan Li}
\author{Chao Tang}
\author{Yong Meng Sua}
\author{Yu-Ping Huang}
 \email{yuping.huang@stevens.edu}
\affiliation{%
 Physics Department, Stevens Institute of Technology, Castle Point at Hudson, Hoboken, New Jersey, 07030, USA\\
 Center for Quantum Science and Engineering, Stevens Institute of Technology, Castle Point at Hudson, Hoboken, New Jersey, 07030, USA
}%

\date{\today}

\begin{abstract}
Wavelength transduction of single-photon signals is indispensable to networked quantum applications, particularly those incorporating quantum memories. Lithium niobate nanophotonic devices have demonstrated favorable linear, nonlinear, and electro-optical properties to deliver this crucial function while offering superiror efficiency, integrability, and scalability. Yet, their quantum noise level--an crucial metric for any single-photon based application--has yet to be understood. In this work, we report the first study with the focus on telecom to near-visible conversion driven by a telecom pump of small detuning, for practical considerations in distributed quantum processing over fiber networks. Our results find the noise level to be on the order of $10^{-4}$ photons per time-frequency mode for high conversion, allowing faithful pulsed operations. Through carefully analyzing the origins of such noise and each's dependence on the pump power and wavelength detuning, we have also identified a formula for noise suppression to $10^{-5}$ photons per mode. Our results assert a viable, low-cost, and modular approach to networked quantum processing and beyond using lithium niobate nanophotonics. 

\end{abstract}

\maketitle


\section{\label{sec:level1}Introduction}

Frequency conversion of single photons across spectral bands is desirable for at least two reasons. The first stems from the anticipated construction of hybrid quantum systems connecting disparate nodes, like cold atoms \cite{Vernaz-Gris2018,Wang2019} and superconducting Josephson junctions \cite{GU20171}, where the photons from each source must be transduced into indistinguishable quantum states. The second has to do with the up-conversion detection of near-infrared single photons for low cost, high detection efficiency, and reduced detector dark counts \cite{Pelc:11,Shentu:13}. For both applications, it is crucial that the quantum noise generated during the conversion process does not degrade the quantum peculiarity of the single-photon signals. This requires not just quantifying the noise level but also identifying its origin to design proper suppression measures. 

Across various platforms, second order nonlinear ($\chi^2$) waveguides are widely deployed for quantum frequency conversion (QFC) for high robustness and care free operations \cite{Shentu:13,Pelc:11,Kuo:13,Maring:18,PhysRevApplied.9.064031}. 
In particular, reverse proton exchange (RPE) or ridge waveguides in periodically poled lithium niobate (PPLN) have been most commercialized owing to their stable recipe, little loss, low noise, and wavelength agility. In those devices, there are two dominant noise processes detrimental to QFC: the spontaneous Raman scattering (SRS) and spontaneous parametric down conversion (SPDC) of the pump \cite{Pelc:11}. SRS is a result of the coupling between the incident pump photons and phonons. SPDC noise is due to non-phase-matched parametric fluorescence that is enhanced by the disorders in poling domain arises from fabrication error during periodic poling\cite{Pelc:10,Pelc:10}. To suppress the noise, far and red detuned pump lasers have been used to suppress these two processes \cite{Pelc:11,Shentu:13,DongAPL}. Yet, for telecom-band quantum signals, their wavelength must be in the vicinity of 2 microns for effective elimination of broadband SPDC noise, rendering the QFC system to be costly, inconvenient, and/or complicated. It also requires a lossy two-stage conversion of telecom quantum signals to match distinct absorption lines of atomic quantum memories.   

Recently, nanophotonic PPLN waveguides have been realized using thin film lithium niobate (TFLN), promising ultrahigh normalized nonlinear efficiency because of much tighter mode confinement \cite{Wang:18, Chen:19, PhysRevLett.125.263602}. Also, they allow the dense integration of a diverse variety of functional components, like electro-optic modulators \cite{Wang2018, Jin:19}, heralded single photon sources \cite{PhysRevLett.124.163603}, and photon detectors \cite{chengAPL2020}, on a single chip. It is thus appealing to realize QFC on the same nanophotonic platform for some impactful quantum applications. Yet, comparing with their bulk-optical counterparts, there are two distinct factors that may significantly impact the noise standing. First, during the production of TFLN, its crystalline structure is first destroyed and recovered, thus potentially altering the phonon modes \cite{2012SPIE.8431E..1DH}. Second, there is significant pump mode leakage into the cladding layer surrounding the PPLN nanowaveguide. Thus a detailed study is warranted to onboard integrated TFLN photonics for many quantum applications.  

In this letter, we report the characterization and a detailed analysis of the quantum noise generated during frequency conversion in a X-cut PPLN nanowaveguide. Specifically, we focus on QFC of telecom C-band to near visible regime around 780 nm. Rather than using a far-detuned pump, we drive the QFC using a slightly detuned pump in observation of the need to transduce single photon in telecom c-band to atomic memories near 780 nm. Additionally, such a small-detuned pumping scheme offers several practical advantages despite notorious challenges in suppressing the noise \cite{Pelc:11,Pelc:11b}. First, it greatly reduces the loss of on/off-chip coupling for both the signal and pump. Moreover it permits good mode overlap in the nanowaveguide, as critical to achieve high QFC efficiency. Third, telecom lasers are pervasively available, inexpensive, and system-integration friendly. Perhaps most importantly, in fiber optical networks such a pump can be derived from the same master laser as the signal, allowing easy and jitter-free synchronization for the whole system, as important for a multitude of distributed applications.   

Here, we measure the total noise level as a function of the applied pump power, at a pulse repetition rate of 10 MHz. Our result finds the total noise level to be $\sim 10^{-4}$ per time-frequency mode in the sum frequency (SF) band when the internal conversion efficiency is about $49\%$ (75\% if there were no propagation loss). We quantitatively analyze the quantum noise created by four different processes in the nanowaveguide, including SRS into the original signal band, SPDC by the residual second-harmonic (SH) of the pump, SRS by the same, and four-wave mixing, as well as those by SRS in the input fiber. Based on our measurement, we outline the means for noise suppression. Also, we measure the Raman spectrum in the small detuning regime, which shows a peak profile similar to their bulk counterparts \cite{Pelc:11, PhysRevB.91.224302}. Together, our results assert the feasibility of using TFLN nanohpotonics for quantum applications, particularly those of distributed quantum processing relying on precise pulse synchronization.

\section{Experimental Setup}
In our experiment, an 8-mm long PPLN nanowaveguide is fabricated on a 500 nm thick, X-cut, magnesium-doped lithium niobate thin film bonded on silicon dioxide that resides on a silicon base. The nanowaveguide is fully etched with a sidewall angle of 67 degree. The top width is measured to be 1.8 $\mu$m. The three interacting waves, pump, signal, and sum-frequency, are  all in the fundamental TE modes to interact through $d_{33}$, the highest nonlinear susceptibility in lithium niobate. The nanowaveguide is periodically poled with a period of 2.5 $\mu$m to compensate for phase mismatching. An inverse taper is fabricated on the input port for a better coupling efficiency of telecom light. More details of the fabrication process are described in Ref.~\cite{Chen:19}. The chip is placed on a stage whose temperature is stabilized at 48$\pm$0.1 degree. To measure the phase-matching curve of the sum frequency generation, the pump wavelength is parked at 1559.7 nm while the signal wavelength is swept. The resulting sum-frequency power over a sweeping bandwidth of 4.5 nm is presented in Fig.~\ref{Images: Fig_1}(a). Here, the baseline comes from the residual second-harmonic generation (SHG) of the pump, due to imperfect poling.

Next, the signal is fixed at the peak sum frequency generation (SFG) wavelength and a power of 214 $\mu$W, while the pump power is gradually increased from 62 mW to 1230 mW. At high pump power, the signal wavelength is tuned slightly around 1540.5 nm to compensate for the thermal effect. The up-converted light is passed though two band-pass filters to reject the residual SH light, before measured by a silicon power meter. The result is shown in Fig.~\ref{Images: Fig_1}(b), which is best fitted with \cite{Roussev:04},
\begin{equation}
    \eta=\eta_\mathrm{max} \mathrm{sin}^2(\sqrt{\eta_\mathrm{nor}P}L).
\end{equation} 
Here, $\eta$ is the internal conversion efficiency, $\eta_\mathrm{max}$ is the maximum conversion efficiency limited by the propagation loss, $P$ is the pump peak power on chip, and $L$ denotes the nnaowaveguide length. Due to the loss, the highest internal conversion efficiency is 65$\%$ when the pump power on chip is 430 mW. Thus, the normalized efficiency is $\eta_{nor}$=863$\%$ $W^{-1}$cm$^{-2}$, thanks to the tight mode confinement and lithium niobate's largest nonlinear susceptibility $d_{33}$. For $L=0.8$ cm, and $P=430$ mW, the conversion efficiency would have been 99.9\% if there were no waveguide loss, i.e., $\eta_\mathrm{max}=1$. 

\subsection{Noise Generated in the Signal Band}
\label{Section NGSB}

\begin{figure}[htbp]
\centering
\fbox{\includegraphics[width=\linewidth]{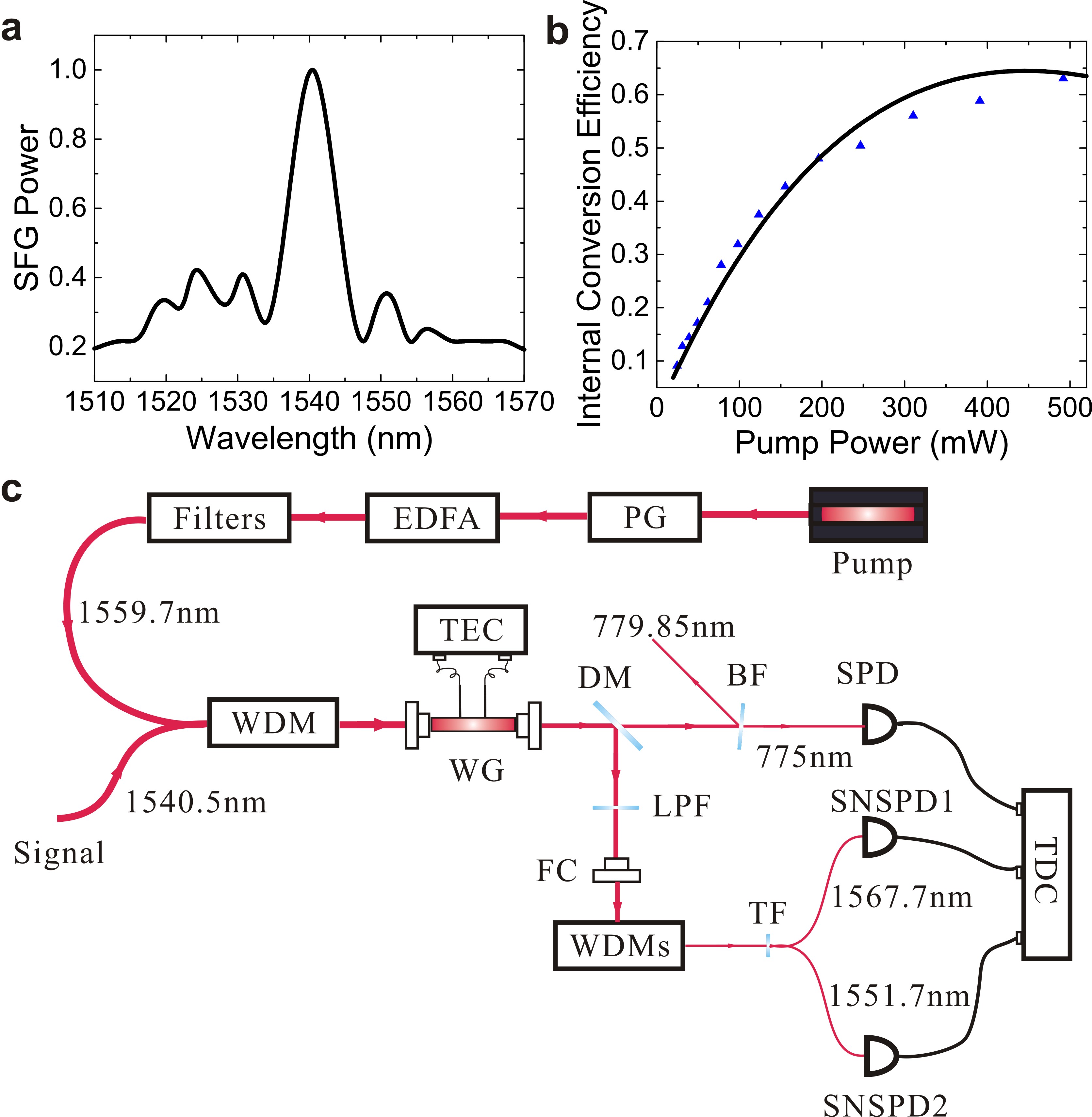}}
\caption{(a) SFG Phase-matching curve. (b) Internal conversion efficiency, where triangles are the experimental data and the black line is the fitted result. (c) Experimental scheme for the noise measurement. PG, pulse generator; WG, waveguide, TEC, thermoelectric cooler; DM: dichroic mirror, BF: band-pass filter, TF, tunable filter; LPF, long pass filter; SNSPD, superconducting nano-wire single photon detector; SPD, silicon single photon detector; TDC, time-to-digital converter.}
\label{Images: Fig_1}
\end{figure}

The experimental setup to measure the quantum noise is shown in Fig.~\ref{Images: Fig_1}(c). The pump is created by carving the output of a 1559.7 nm continuous-wave laser into a 10 MHz, 10 ns pulse train by an electro-optical modulator. Then, the pulses are fed into an erbium-doped fiber amplifier(EDFA) with three 100 GHz wavelength division multiplexers (WDM's) at the output to suppress the spontaneous emission noise of the EDFA. The signal and pump are combined through a WDM, and injected into the nanowaveguide with a lensed fiber (OZ optics). The output is collected by another lensed fiber and collimated into free space through a fiber coupler. The telecom and SF light are separated by a short-pass dichroic mirror. The SF light is further cleaned by two band-pass filters with a 3-nm bandwidth to reject the residual SH light at 779.9 nm by over 120 dB. Then the up-converted signal at 775 nm and the remaining noise are recorded by a visible single-photon detector (ID100), with a detection efficiency of 15$\%$. Meanwhile, the telecom light is directed to a fiber coupler after passing through a long-pass filter. Three identical WDM's centered at 1559.7 nm are serialized to dump the pump power at their rejection ports. Next, one WDM at 1551.7nm is followed by two tunable filters(OTF-910) connected to its pass and reflection ports, respectively. The bandwidth of two filters are measured to be 0.6 nm and 0.8 nm respectively. In our experiment, we don't directly measure noise at signal wavelength 1540.5 nm directly to avoid potential underestimation of the noise counts \cite{Maring:18}. Instead, the wavelength of those two tunable filters are set to be 1551.7nm and 1567.7nm, respectively, symmetrically residing at two sides of the pump and both out of the up-conversion window.  With this scheme, SPDC noise from the residual SH light is quantified. Photons from the two tunable filters are recorded by a four-channel superconducting nanowire single-photon detector (SNSPD, IDQ 281), whose detection efficiency is around $85$\%. In this way, the noise scattered into the signal and SF wavelength bandwidth can be independently analyzed at the same time.

There are four possible mechanisms to create background photons at the signal wavelength: the SHG-SPDC noise, SRS in both the input fiber and the nanowaveguide, and noise through spontaneous four-wave mixing (SFWM). As the pump is on the Stokes side of the signal, there is no SPDC noise generated directly by the pump. Also, SFWM noise created in the four-meter input fiber is negligible \cite{Li:04}. Our goal is to quantify their individual contributions through their distinct dependencies on the pump power. 

\begin{figure}[htbp]
\centering
\fbox{\includegraphics[width= \linewidth]{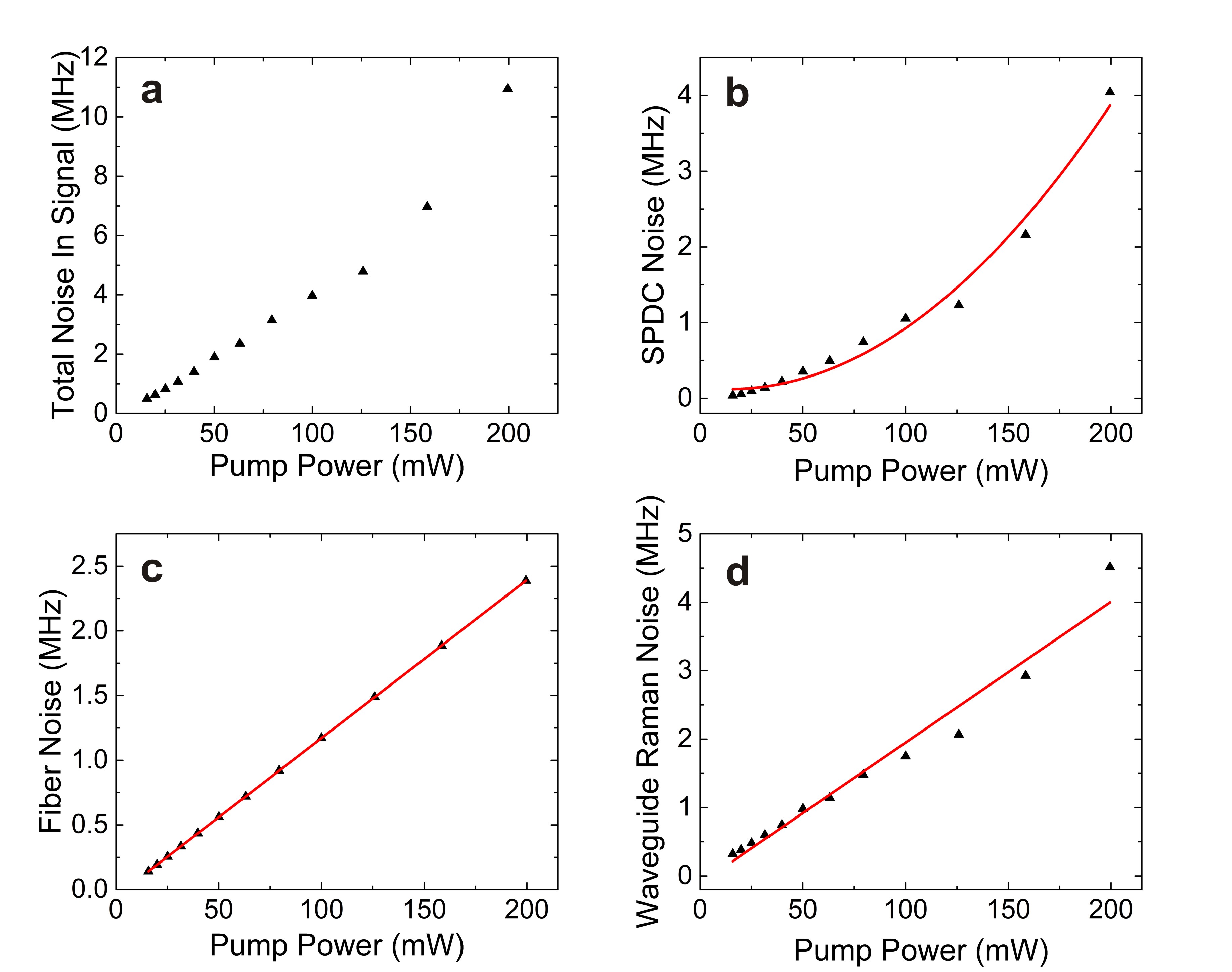}}
\caption{Noise in the signal band as a function of the pump power. (a) The total on-chip count rate over a detection bandwidth of 0.6 nm. (b) SPDC count rate versus the pump power. (c) In-fiber Raman counts. (d) Waveguide Raman count rate by subtracting SPDC and fiber noise from the total noise. In all figures, the triangles are the experimental data and red lines are quadratic (in b) and linear (in c and d) fitting results.}
\label{Images: telecbandnoise}
\end{figure}

Among them, both the SHG-SPDC and SFWM processes inject noise to the frequency upconversion by photon pair generation (PPG). The first involves a cascaded $\chi^{(2)}$ process, where two pump photons merge into a visible photon that spontaneously down-converts a pair of nondegenerate photons over a broad spectrum. In the latter case, photon pairs are directly generated from the pump through the Kerr nonlinearity. Thus both occur at a rate quadratic to the applied pump power. To quantify the two, the signal input is disconnected, and  the noise photons at 1551.7 nm and 1567.7 nm, as well as their coincidence, are registered by a time to digital converter (ID-900), as the pump power increases. The total coupling loss from the input of the nanowaveguide to the input fiber of the SNSPD are $20.9$ dB for 1551.7 nm and $19.4$ dB for 1567.7 nm, respectively. The input coupling loss of the nanowaveguide is estimated to be 4 dB for both wavelengths, according to the characterization of a similar nanowaveguide with two identical inverse tappers of the same parameters. The transmission efficiency of the noise photons, from the nanowaveguide to the detectors, are $\eta_1$=$1.74\%$ and $\eta_2$=$2.45\%$ for the two wavelengths , respectively, after taking into account the detector's efficiency. The number of total noise photons at the chip output $R_\mathrm{total}$ and the PPG on chip $R_\mathrm{ppg}$ are simply
\begin{align}
 R_\mathrm{total}= \frac{R_{1551.7}}{\eta_1}, \label{spdc} \\
 R_\mathrm{ppg}= \frac{R_\mathrm{c}}{\eta_1 \cdot \eta_2}, \label{ppg}
\end{align}
where the second equation assumes PPG to contribute dominantly to the coincident counts. $R_{1551.7}$ is the noise count 1551.7 nm, and $R_\mathrm{c}$ is the coincidence counting rate. The results are presented in Fig.~\ref{Images: telecbandnoise}(a) and (b). In the first figure, the total noise increases nearly linearly with low pump power, but showing a quadratic dependency for high pump power. In the second figure, the PPG rate is fitted well with a quadratic function. Together, they suggest that the noise from PPG becomes dominating in the high pump power regime. In fact, for pump power $\sim$200 mW, corresponding to an internal up-conversion efficiency close to 50$\%$, the PPG noise accounts for 42$\%$ of the total noise. 

In the above estimation, we have assumed a flat SPDC gain spectrum over a bandwidth of 11 nm, from 1551.7 nm to 1540.5 nm, which is typically the case \cite{Lu:19, Elkus:19}. Also, Eq.~(\ref{ppg}) is valid only when PPG dominates other noise sources. To verify that, we monitor the coincidence to accident ratio (CAR) for those power settings. Our result show that as the pump power increases, CAR drops from 147 to 35, which nonetheless remains significantly higher than 1. It confirms that the coincident counts mostly come from PPG and Eq.~(\ref{ppg}) is valid. 

To quantify the SFWM's contribution to the total noise level, another nanowaveguide without poling, of the similar geometry but with a 100 nm wider top, is fabricated on the same chip. Its CAR is measured using the same pump. Even at the highest power setting of the EDFA for the pump, the CAR turns out to be 1. This result rules out SFWM's contribution to the PPG. 

Another noise source is the spontaneous Raman scattering in the input fiber after the WDM combiner. The Raman noise after the nanowaveguide, on the other hand, is negligible as the pump power is greatly reduced passing through the chip. To quantify it, a 14.5 dB in-line attenuator is inserted in the place the chip and its coupling optics to emulate its total loss, while all other parts are untouched. The in-fiber Raman photon counts at the end of the 4 meters fiber along the in-fiber pump power are obtained by accounting for the transmission loss and detection efficiency. Subsequently, they are curve fitted with a linear relationship. The on chip counts as a function of the on chip pump power are plotted in Fig.~\ref{Images: telecbandnoise}(c), after including 4 dB coupling loss to the nanowaveguide. From the result, the in-fiber Raman noise takes up a significant part of the total noise, around 30$\%$ when the internal conversion efficiency is less than 40$\%$. It can be reduced by using a shorter input fiber in the future. 

Lastly, the Raman noise created in the nanowaveguide is obtained after subtracting the SPDC noise counts and the in-fiber Raman counts from the total noise counts. The result is shown in Fig.~\ref{Images: telecbandnoise}(d), which exhibits a good linear dependence on the pump power, as expected. It takes the lead before the on chip pump power reaches 125 mW. For higher power, the SPDC noise becomes the main source, owning to its quadratic dependence on the pump power. 

\subsection{Noise Generated in the SF Band}
The above sources result in the noise in the telecom band. Yet, only those falling in the phase matching window can be converted and appear as the noise at the wavelength of the upconverted signal. Another noise, which occurs concurrently, is the direct SRS of the residual SH light at 779.85 nm.

\begin{figure}[htbp]
\centering
\fbox{\includegraphics[width=\linewidth]{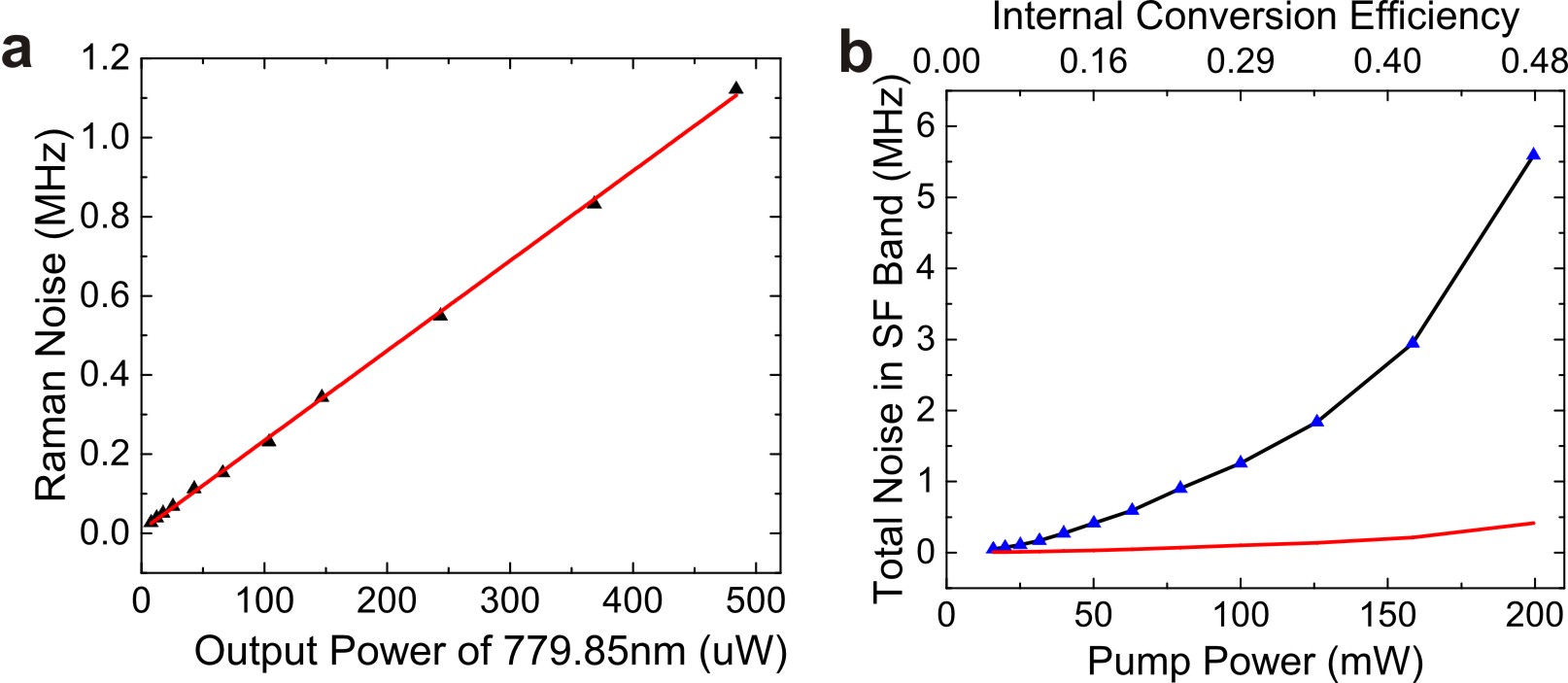}}
\caption{Noise analysis in the SF band. (a) The on chip Raman noise count rate $R_\mathrm{v}$, with a bandwidth of 3nm, along the output power of 779.85 nm light. The red line is the linear fitting result, and triangles are experimental data. (b) The total noise in the SF wavelength. Red line is the deduced Raman noise due to the residual SH power along the pump power at 1559.7 nm. Triangles are the total noise in the SF wavelength.}
\label{Images: visiblebandnoise}
\end{figure}

To accurately quantify the noise level through this SHG-SRS process, it is necessary to eliminate the above signal band noise in Section \ref{Section NGSB}, as it can be upconverted to the SF band. To this end, we directly apply pulses at 779.85 nm, the second harmonic of the pump, to the nanowaveguide and measure the resulting noise counts. The SH pulses are created by SHG in a separate PPLN ridge waveguide (HC Photonics), and cleaned by three band-pass filters before coupled into the nanowaveguide. To ensure that the input SH light resides in the fundamental TE mode of the waveguide, it is tested against a difference frequency generation (DFG) process with a 1540 nm laser to create 1579.86 nm light. The polarization of the SF light is tuned to maximize the output power of 1579.86 nm light. Afterwards, the 1540 nm light will be removed when we collect the Raman noise counts. The results are shown in Fig.~\ref{Images: visiblebandnoise}(a), where the counting rate $R_\mathrm{v}$ is a good linear function of the SH power $P_\mathrm{SH}$. Note that this is an overestimate of the direct SRS level. This is because in this experiment, the SH light power is nearly constant across the nanowaveguide, so that $R_\mathrm{v}=\xi_\mathrm{R} P_\mathrm{S} L$, with $\xi_\mathrm{R}$ the SRS efficiency. During the upconversion, however, the residual SH light is created along the waveguide with a quadratically increasing power. Thus the Raman noise rate due to the residual SHG light scattering could be calculated as $R_t=\eta_{o} P^2_\mathrm{p} \int_{0}^{L} \xi_\mathrm{SHG}^2\xi_\mathrm{R}  x^2 dx=\frac{1}{3} \eta_{o} P^2_\mathrm{p} \xi_\mathrm{SHG}^2 \xi_\mathrm{R} L^3$, where $\eta_{o}$ is the output coupling efficiency. $P_\mathrm{p}$ is the input pump power at 1559.7 nm that is assumed undepleted through the nanowaveguide. $\xi_\mathrm{R}$ and $\xi_\mathrm{SHG}$ denote the SHG and SRS efficiencies, respectively. As the generated SH power (779.85 nm) at the output $P_\mathrm{S}=\eta_{o} P^2_\mathrm{p} \xi_\mathrm{SHG}^2 L^2$, we have 
\begin{equation}
\label{rtin}
R_t=\xi_\mathrm{R} P_\mathrm{S} L/3.
\end{equation}
Therefore, for the same SH power at the output,
\begin{equation}
\label{eq:one-third}
    R_\mathrm{t}=R_\mathrm{v}/3, 
\end{equation}


In Fig.~\ref{Images: visiblebandnoise}(b), we plot the total noise in the SFG as a function of the applied pump power at 1559.7 nm. Comparing with Fig.~\ref{Images: visiblebandnoise}(a) and using Eq.~\ref{eq:one-third}, the direct SRS followed by the residual SHG contributes at most 11$\%$ of the total noise.

\subsection{Raman spectrum}
From the above measurement results, the indirect SRS noise---by the pump first Raman scattered into the signal band and subsequently upconverted---adds significantly to the total noise in the SF band. Hence, it is helpful to measure the SRS spectrum in the neighborhood of the signal and subsequentially choose the pump wavelength at one of its valleys, a method first adopted in SFWM in optical fibers \cite{Li:05}.

\begin{figure}[htbp]
\centering
\fbox{\includegraphics[width=0.8\linewidth]{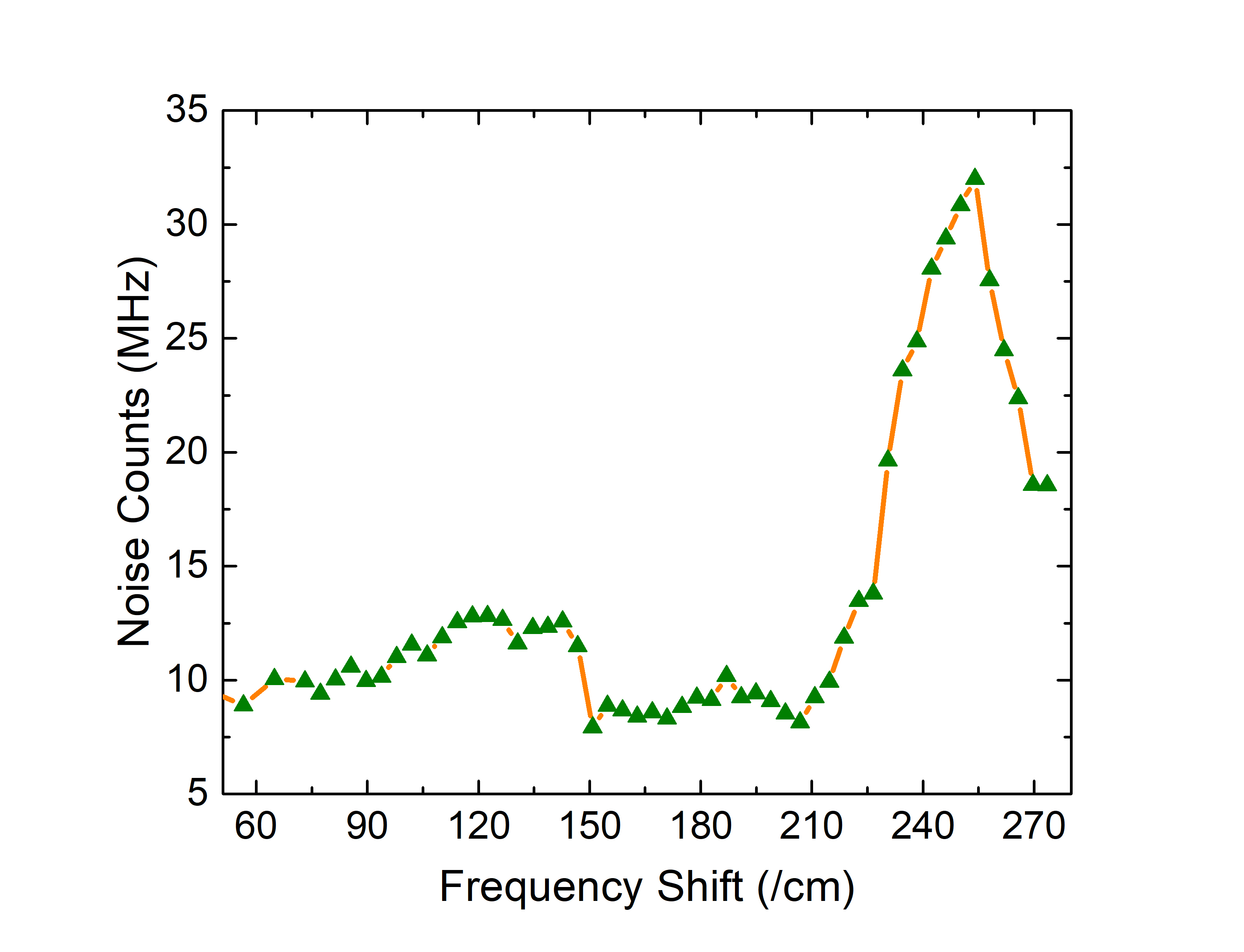}}
\caption{SRS spectrum in a small detuning regime by noise photon counting.}
\label{Images: Ramanspectrum}
\end{figure}

In our experiment, a pulse train with 10 MHz repetition rate and 10 ns pulse width is prepared at 1537.5nm, to use as the pump. It is amplified by an EDFA, followed by two 100-GHz WDMs to suppress the side band noise at the output. The resulting pump, with 25 dBm peak power, is coupled into the nanowaveguide. To prevent the pump photon from reaching the SNSPD, we first dump the pump with two WDMs whose through band is centered at 1537.5 nm. The output is further filtered by two tunable filbers in series (bandwidth: 0.6 nm and 0.8 nm), to provide over 220 dB rejection of the pump photons. Photon counts are taken as the tunable filters are swept from 1550 nm to 1605 nm at step size of 1 nm. The noise spectrum on the Stokes side is presented in Fig.~\ref{Images: Ramanspectrum}. An obvious peak is erected at 254/cm, which coincides the previous Raman spectrum study in congruent lithium niobate\cite{Pelc:11, PhysRevB.91.224302}. The result also suggests a suitable region free of Raman active peak, from 150$/cm$ to 210$/cm$, for photon conversion by small-detuned pumping with relative low noise. 

\section{Noise Management Scheme}
Lastly, we summarize the noise performance of the PPLN nanowaveguide and discuss potential ways for suppression. Table I lists the total noise counts per time-frequency mode in the SF band and their attributions, for the current experimental results. Also listed are the estimates after a variety of modifications to the system configurations. Here, a time-frequency mode is defined a minimum uncertainty mode with the least area in the time and frequency phase space, as defined by the normal-mode description of time-bandwidth limited detection \cite{Bell1961,PhysRevA.42.6794,PhysRevA.73.043807}. For each pump pulse, the number of time-frequency modes is given by $BT$, where $B$ ($2\pi B$ is the corresponding angular frequency) is the area subtended by the (effective) spectral filter profile and $T$ is the effective width of the time window. 

In our case, the noise photons in the signal band are counted after passing through a 0.6 nm filer, so that $B=75$ GHz. With $T=10$ ns, the number of detection modes is 750. The SHG-SRS noise is measured with a 3 nm filter, so that the mode number is $1.5\times 10^{4}$. When counting the total noise photons, although the same 3 nm filter is used in the SF band, the phase matching bandwidth of the PPLN nanowaveguide is only 4.5 nm in the telecom band, corresponding to 1.1 nm in the visible band. Thus only those noise photons created first within the 4.5 nm band around the signal wavelength can be upconverted to the SF band. Taking this into account, and that SHG-SRS contributes only marginally to the total noise level, the detection mode number is calculated with $B=0.56$ THz to be 5600.

\begin{table*}[h]
\newcommand{\tabincell}[2]{\begin{tabular}{@{}#1@{}}#2\end{tabular}}
	\centering
	\begin{tabular}{|c|c|c|c|c|c|}
		\hline
		System Configuration & \tabincell{c}{In-waveguide SRS, \\ Signal Band}  & \tabincell{c}{In-fiber SRS, \\ Signal Band} & \tabincell{c}{SHG-SPDC, \\ Signal Band} & \tabincell{c}{SHG-SRS, \\SF Band}& \tabincell{c}{Total, \\ SF Band  \\} \\
		\hline
		Current Experiment, 49\% Conversion (75\% if no loss) & 60.26 & 31.87 & 54.00& 0.27 & 9.96\\
		\hline
		Input Fiber Reduced to 1 m & 60.26 & 7.97 & 54.00 & 0.27 & 6.15\\
		\hline
		Input Fiber Eliminated & 60.26 & 0 & 54.00 & 0.27 & 4.88 \\
		\hline
	    Residual SHG Reduced by Half & 60.26 & 31.87& 27.00 & 0.14 & 9.39\\
		\hline
	    Residual SHG Eliminated & 60.26 & 31.87 & 0 & 0 &8.81 \\
		\hline
		Reduced Input Fiber and Residual SHG & 60.26 & 7.97 &27.00 &0.14 &5.57 \\
		\hline
		No Input Fiber or Residual SHG &  60.26 & 0 &0 &0 &3.73\\
		\hline
	\end{tabular}
	\caption{The total noise photon counts under different configurations and their attributions. All numbers are given in $10^{-5}$ photons per detection mode. Note that because the noise in the signal band is created along the waveguide, only a fraction is converted into the sum frequency (SF) band even the conversion efficiency is high for the input signal. 
	The residual SHG is reduced by improving the uniformity of the periodic poling.}
	\label{t2}
\end{table*}

As seen in Table \ref{t2}, in the current configuration, the total noise created in the SF band is $ 10^{-4}$ per mode. By reducing the input fiber to 1 meter and improving periodic poling to suppress residual SHG, it can be lowered to $5.57\times 10^{-5}$. As shown in the same table, this noise can be further suppressed to $3.73\times 10^{-5}$ by totally eliminating the input fiber and residual SHG. 

Ultimately, the remaining yet most critical noise source compromising the single-photon conversion in PPLN nanowaveguides is the in-waveguide SRS that produces noise photon in the phase-matching bandwidth of SFG. SRS noise can be reduced by lowering the operating temperature of the PPLN nanowaveguides\cite{Kuo:18} due to the temperature-dependent nature of Raman scattering \cite{PhysRev.168.1045}. A strategy one can adopt is designing the period of PPLN nanowaveguides for operating at much lower temperature by taking advantage of the tunability of phase matching wavelength\cite{Chen:20,Luo:18}. This will suppress the in-waveguide SRS, potentially leads to several-fold decrease of intrinsic noise photon\cite{Kuo:18} after discounting the input fiber and residual SHG.

\section{Conclusion}
We have reported the first dedicated experimental characterization of the quantum noise and its attributions in a nanophotonic waveguide on thin-film lithium niobate. While a large, red detuned pump is usually required to suppress quantum noise during frequency conversion, our results  reveal that, by using a small detuned pump in the same telecom band, the noise level may be acceptable for some quantum applications. Among multiple noise sources, the spontaneous Raman scattering in the waveguide and SPDC following the residual SHG of the pump dominate. The former can be mitigated by parking the pump wavelength at the valley of the Raman gain spectrum and lowering the operating temperature of the PPLN nanowaveguide. The latter can be suppressed by reducing the residual SHG by, for example, improving the periodic poling quality. Our results assert the prospect of creating highly integrated nanophotonic devices on thin-film lithium niobate for quantum optics tasks where scalablity and modularity are desirable.


\begin{acknowledgments}
This research is supported in part by National Science Foundation (Awards: 1806523, 1842680). Device fabrication is performed at at the Advanced Science Research Center NanoFabrication Facility at the Graduate Center of the City University of New York.
\end{acknowledgments}

\nocite{*}

\bibliography{apssamp}

\end{document}